\documentclass[prl,twocolumn,aps,superscriptaddress,footinbib]{revtex4-2}
\usepackage{xspace,amsmath,amsfonts,amssymb,amsbsy}
\usepackage{ntheorem}
\usepackage{graphicx,epstopdf}
\usepackage{bbm}
\usepackage{footnote}
\usepackage{braket}
\usepackage[usenames,dvipsnames]{xcolor}
\usepackage[breaklinks=true]{hyperref}
\hypersetup{
colorlinks   = true, 
urlcolor     = blue, 
linkcolor    = blue, 
citecolor    = red 
}
\usepackage{soul}
\usepackage{ulem}
\usepackage{float}

\makeatletter

\begin{document}
\title{Collective three-body interactions enable a robust quantum speedup}

\author{Haoqing Zhang}
\affiliation{JILA, NIST and Department of Physics, University of Colorado, Boulder, Colorado 80309, USA}
\affiliation{Center for Theory of Quantum Matter, University of Colorado, Boulder, Colorado 80309, USA}
\author{Anjun Chu}
\affiliation{JILA, NIST and Department of Physics, University of Colorado, Boulder, Colorado 80309, USA}
\affiliation{Center for Theory of Quantum Matter, University of Colorado, Boulder, Colorado 80309, USA}
\author{Chengyi Luo}
\affiliation{JILA, NIST and Department of Physics, University of Colorado, Boulder, Colorado 80309, USA}
\author{Chitose Maruko}
\affiliation{JILA, NIST and Department of Physics, University of Colorado, Boulder, Colorado 80309, USA}
\author{Eliot A.~Bohr}
\affiliation{JILA, NIST and Department of Physics, University of Colorado, Boulder, Colorado 80309, USA}
\author{James K. Thompson}
\affiliation{JILA, NIST and Department of Physics, University of Colorado, Boulder, Colorado 80309, USA}
\author{Ana Maria Rey}
\affiliation{JILA, NIST and Department of Physics, University of Colorado, Boulder, Colorado 80309, USA}
\affiliation{Center for Theory of Quantum Matter, University of Colorado, Boulder, Colorado 80309, USA}

\date{\today}
\begin{abstract}
We show that collective three-body interactions (3BIs), implementable with $N$ atoms loaded inside an optical cavity, offer a significant advantage for preparing complex multipartite entangled states.  Firstly,  they enable a speedup of order $\sim N$ in preparing generalized Greenberger-Horne-Zeilinger (GHZ) states, outperforming conventional methods based on all-to-all two-body Ising interactions.  Secondly, they saturate the  Heisenberg bound in phase estimation tasks using a time-reversal protocol realized through simple rotations and followed by experimentally accessible collective spin measurements.
Lastly, compared with two-body interactions (2BIs), in the presence of cavity losses and single particle decoherence, 3BIs feature a high gain in sensitivity  for moderate atom numbers and in large ensembles a fast entanglement generation  despite constraints in parameter regimes where they are implementable.
\end{abstract}
\maketitle

\textit{Introduction--} 
A central goal in quantum science is the ability to engineer and control microscopic interactions among particles. Tunable two-body interactions (2BIs) have long served as the foundation of this effort, underpinning entanglement generation, spin-exchange dynamics, and universal quantum gates. Recent advances, however, have pushed beyond this paradigm, enabling the controlled realization of genuine multi-body interactions in which three or more particles interact simultaneously. These higher-order couplings have been proposed and realized across diverse platforms, including superconducting qubits~\cite{fedorov2012implementation,kim2022high,eriksson2024universal}, trapped ions~\cite{katz2023demonstration,băzăvan2024squeezing,saner2024generating}, ultracold atomic gases~\cite{dai2017four,Goban2018,Will2010,kraemer2006evidence,Hadzibabic20173bodycontact}, and Rydberg atom arrays~\cite{levine2019parallel,AhnRydberg2020,kim2024realization}. Beyond their fundamental interest, multi-body interactions offer powerful new routes for accelerating quantum gates~\cite{nielsen2010quantum} and for simulating complex many-body Hamiltonians relevant to chemistry, condensed matter, and high-energy physics~\cite{threebodyRMP2013,buchler2007three,BernardPSpin1980,Berg2009,Fujita1957,Pinto2024}.

Realizing such couplings, however, remains challenging: (i) They typically arise from higher-order processes, leading to slower dynamics and enhanced decoherence; (ii) Genuine multi-body effects are difficult to isolate from dominant lower-order terms; and (iii) previous demonstrations have been limited to few-particle systems.
A recent experiment \cite{luo2025realization} addressed these issues by engineering collective three-body interactions (3BIs) in an ensemble of $N\sim10^3$ cold atoms coupled to an optical cavity, using momentum states as pseudo-spins. Here, 3BIs emerge from a cavity-mediated six-photon process -- higher orders processes compared to the four-photon process producing two-body couplings. Although the bare three-body coupling is weaker, the cavity’s all-to-all connectivity enhances collective 3BIs to comparable strength with collective 2BIs~\cite{luo2025hamiltonian}, suggesting faster dynamics in larger arrays. But to date, observations have been restricted to mean-field timescales, leaving the role of 3BIs in genuine quantum dynamics an open question.

In this letter, we present a detailed analysis of the beyond MF dynamics induced by collective 3BIs and demonstrate they indeed offer a significant advantage for sensing. 
Due to their intrinsic $Z_3$ rotational symmetry, generalized Greenberger-Horne-Zeilinger (GHZ) states can be prepared with high fidelity for specific atom numbers, at a speed that is $\sim N$ faster compared to the conventional one-axis twisting (OAT) scheme based on all-to-all two-body Ising interactions.
For generic particle numbers, we also observe a rapid growth of multi-particle correlations and quantum Fisher information (QFI), reaching the Heisenberg scaling (HS) desired for phase estimation within experimentally realistic time. Such gain can be extracted using a time-reversal (TR) protocol achievable via simple single-particle rotations~\cite{garttner2017measuring,garttner2018relating,lewis2019unifying,chu2021quantum}, and measurements of collective single-particle observables.

Even more importantly, we demonstrate that the dynamics remain robust against realistic dissipation inherent to cavity-QED systems. Whereas most previous work on entanglement generation focuses on scalability, our results highlight a complementary advantage: for moderate atom numbers, $N$, and cooperativity $C$, collective 3BIs produce a marked enhancement in metrological gain relative to generic 2BI models, especially since these 2BI models exhibit strong deviations from analytic scaling at limited cooperativity, $NC < 100$~\cite{koppenhofer2023revisiting}. For larger ensembles, 3BIs further enable faster preparation of target entangled states, an equally critical factor given that coherence times are constrained by technical noise and dissipation.

\begin{figure}[!t]
\centering
\includegraphics[width=0.48\textwidth]{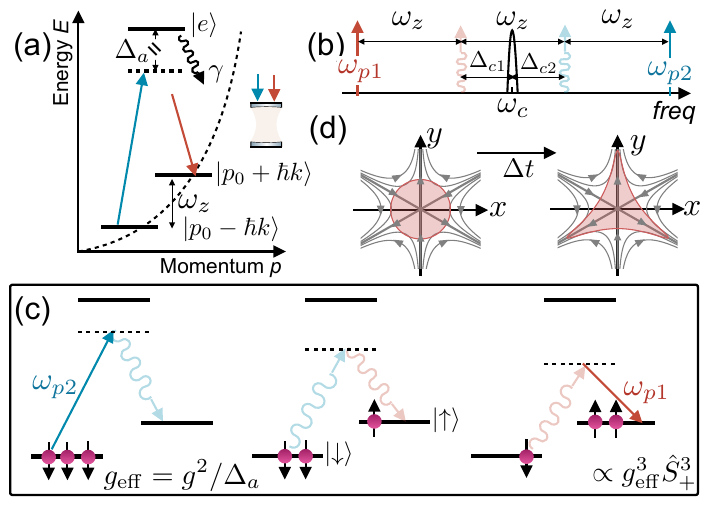}
\caption{(a) Energy level diagram of the atomic ensemble used to engineer the 3BIs, where a pair of momentum states, $\ket{p_0\pm\hbar k}$, serves as the pseudo-spins $\ket{\uparrow/\downarrow}$~\cite{luo2025realization}. Inset: schematics of the setup. An optical cavity is driven by two dressing lasers with frequencies  $\omega_{p1,2}$ (red and blue arrows).
(b) Corresponding Frequency diagram.
(c) Two dressing lasers induce a six-photon process that flips three spins collectively between $\ket{\downarrow,\downarrow,\downarrow}$ and $\ket{\uparrow,\uparrow,\uparrow}$ via an intermediate excited state. Virtual cavity photons at two frequencies are shown as red and blue wavy lines (see text).
(d) Mean-field dynamics for an initial spin along $z$. Gray lines show classical trajectories with trifurcation points; bold arrows mark the separatrix.
\label{fig:1}}
\end{figure}

\textit{Model--} We consider the experimental setup recently realized in Ref.~\cite{luo2025realization} and illustrated in Fig.~\ref{fig:1}(a)(b). An ensemble of $N$ cold atoms is confined in a high-finesse optical cavity, where two accessible motional ground states, denoted by $\ket{p_0\pm \hbar k}$ and separated by an energy $\hbar \omega_z$, serve as the pseudo-spin $\ket{\uparrow/\downarrow}$. The cavity mode at frequency $\omega_c$ is detuned by $\Delta_a$ from the optical transition $\ket{\uparrow/\downarrow}$ to $\ket{e}$, with single-photon Rabi frequency $g$ and spontaneous emission rate $\gamma$.
The cavity is driven by two simultaneously applied lasers with frequencies $\omega_{p1}$ and $\omega_{p2}$ (separated by $3\omega_z$).
The detuning $|\Delta_a|$ is assumed to be much larger than all other relevant frequency scales.
Upon adiabatic elimination of both the excited state and the cavity field, an effective collective 3BI emerge~\cite{luo2025hamiltonian}:
\begin{equation}
\hat{H}_3 = \chi_3 (\hat{S}_+^3 + \hat{S}_-^3),
\end{equation}
where $\hat{S}_{\pm}=\hat{S}_{x}\pm i \hat{S}_{y}$ are collective spin raising and lowering operators, and $\hat{S}_{\alpha={x,y,z}}$ denote the components of the total spin operator. The basic idea is illustrated in Fig.~\ref{fig:1}(c). Any group of three atoms in the $\ket{\downarrow}$ state, can flip their pseudo-spin to $\ket{\uparrow}$ and the other way around, in a process which starts when one of the atoms absorbs a photon from the drive with frequency $\omega_{p2}$.  Since the $\ket{e}$ level is  far from resonance, it is only virtually populated and the excited atom quickly emits a cavity photon (blue squiggly arrow) that transfers it to $\ket{\uparrow}$. The emitted photon at frequency $\omega_{p2}-\omega_z$ is also blue detuned from the cavity by $\Delta_{c,2}$, and virtually absorbed by the second atom which quicly transitions to $\ket{\uparrow}$, after emitting a second virtual cavity photon (red squiggly arrow) with frequency $\omega_{p1}+\omega_z$, now red detuned from the cavity by $\Delta_{c,1}$. This last virtual photon is finally absorbed by the third atom, which transitions to $\ket{\uparrow}$ by stimulated emitting a photon to the second pump with frequency $\omega_{p1}$. The net  process by which three atoms in 
$\ket{\downarrow}$, all simultaneously flip to $\ket{\uparrow}$, can be made resonant when the energy gain of the atoms is compensated by the pump photons, $\omega_{p2}-\omega_{p1}=3\omega_z$. The rate at which an atom absorbs/emits a pump photon and subsequently emits a cavity photon is given by the two-photon coupling strength $g_{\rm eff}=g^2/\Delta_a$.

Under the chosen  drive configuration with $\Delta_{c2}=-\Delta_{c1}\equiv \Delta_c$~\cite{luo2025realization,zhang2025solitons}, all 2BIs -- such as the Two-Axis Twisting (TAT) interaction $\hat{H}_{\rm TAT} = \chi_2^{\rm TAT} (\hat{S}_+^2 + \hat{S}_-^2)$~\cite{luo2025hamiltonian} and the exchange interaction $\hat{H}_{\rm ex} = \chi_2^{\rm ex} \hat{S}_+ \hat{S}_-\approx \chi_2^{\rm ex} (\vec{S}\cdot \vec{S}- \hat{S}_z^2)$ (or OAT $\hat{H}_{\rm OAT} = \chi_2^{\rm OAT} \hat{S}_x^2)$, with $\vec{S}\cdot \vec{S}=\sum_{\alpha=x,y,z} \hat{S}_{\alpha}^2$ ~\cite{ThompsonMomentumExchange2024} -- are off-resonant or suppressed by destructive interferences, respectively.

\textit{MF Dynamics--} 
At the MF level, the  Bloch vector $\mathbf{S}(t)\equiv  \{S_x(t),S_y(t) ,S_z(t) \}$ with  $S_\alpha(t)= \langle \hat{S}_\alpha(t) \rangle$, simply evolves under the non-linear set of equations, $\dot{S}_{+}  =-6 i \chi_3 S_z S_-^2$ and   $
\dot{S}_{z}=-3i\chi_3 ( S_+^3- S_-^3)$. The MF flow lines~\cite{luo2025hamiltonian,PoggiDeutsch2023PRXQuantum} can be obtained by initializing the system in a spin coherent state, described by an 
initial  MF Bloch vector $\mathbf{S}(0)\equiv  N/2 \{x, y, \sqrt{1-x^2-y^2}\}$ and then recording the change $\Delta\mathbf{S}\equiv  \mathbf{S}(\Delta t)-\mathbf{S}(0)$ after a short time evolution, $\Delta t$.  In Fig.~\ref{fig:1}(d) we show an example of the flow line map near the north-pole direction on the Bloch sphere.  Under the MF evolution, the north pole is an unstable fixed point and can be identified as a \textit{trifurcation point}. The dynamics along the three separatrix branches lead to attractive (at angles $\varphi = \pi/6, 5\pi/6, 3\pi/2$) or diverging (at $\varphi = \pi/2, 7\pi/6, 11\pi/6$) trajectories, a feature qualitatively distinct from the nonlinear dynamics arising from 2BIs. The  flow-line-map of $\hat{H}_3$ 
was experimentally measured in Ref.~\cite{luo2025hamiltonian}  showing excellent agreement with theoretical predictions.
\begin{figure}[!t]
\centering
\includegraphics[width=0.45\textwidth]{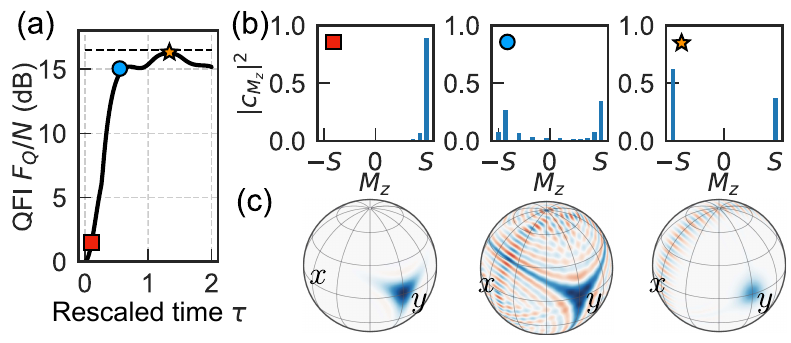}
\caption{(a) Quantum Fisher information (QFI) growth vs time for a system with $N=45$ atoms, starting from an initial spin coherent state aligned along the $z$-axis. The dashed line indicates the Heisenberg limit (HL).
(b) Spin distribution function, $|c_{M_z}|^2$,  in the Dicke basis $\ket{S=N/2, M_z}$, and (c) the corresponding spin Wigner function, after evolution under the three-body Hamiltonian $\hat{H}_3$ at $\tau = 0.1$ (red square), $0.66$ (blue circle), and $1.3$ (yellow star), with rescaled evolution time $\tau=\chi_3 N^{\frac{3}{2}} t$. A final $\pi/2$ rotation about the $x$-axis is applied to facilitate visualization.  
\label{fig:2}}
\end{figure}

\textit{Entanglement generation --}  
The utility of $\hat{H}_3$ for the preparation of metrologically useful entangled states can be quantified by the QFI, which tells the best precision that can be achieved in parameter estimation for a given quantum state across all physically realizable
measurements and estimators. Quantitatively, for a pure state and a unitary parameter encoding, the QFI  can
be expressed in terms of the variance of the operator
responsible for phase encoding.  In our case, the relevant operator is  $\hat{S}_z$,  and thus $F_Q=4 (\langle \hat{S}_z^2\rangle- \langle \hat{S}_z\rangle^2)$. We compute the
QFI as a function of time for an initial state $\ket{\psi_0}$  prepared along the $\hat{z}$ direction, $\ket{\psi_0}=\ket{S=N/2, M_z=N/2}$, which has no MF dynamics.  Here $\ket{S, M_z}$ denote the fully symmetric Dicke states, which are simultaneous eigenstates of the $\vec{\hat S}\cdot \vec{\hat S}$  and $\hat{S}_{z}$ operators, with eigenvalues $S(S+1)$,  $S\in \{0,1,\dots N/2\}$ and $M_z\in \{-S, -S +1,\dots, S\}$ respectively. Since the Hamiltonian  commutes with $\vec{\hat S}\cdot \vec{\hat S}$, $S$ is conserved. In general, the dynamical evolution is given by $\ket{\psi_t} = \sum_{m=0}^{N/3} c_{N/2-3m}(t)\ket{N/2,N/2-3m} $ and thus only every third state in the fully symmetric Dicke ladder is populated.  

In Fig.~\ref{fig:2}(a), we show the time evolution of the QFI for a system of $N=45$ atoms and in Fig.~\ref{fig:2}(c) we show snapshots of the spin Wigner distribution of the state~\cite{supplement}, for three representative time: $\tau = 0.1$ (red square), $0.66$ (blue circle), and $1.3$ (yellow star) with the rescaled time $\tau\equiv\chi_3 N^{\frac{3}{2}} t$. 
At $\tau = 0.1$, the state is already non-Gaussian and exhibits a $Z_3$ rotational symmetry, reflecting the discreteness of the Dicke ladder. 
This structure, analogous to number-phase duality in the bosonic system, could be useful for quantum error correction~\cite{grimsmo2020quantum,valahu2024quantum}.
At $\tau_{\rm opt,3} = 2/3$, when the QFI reaches its first peak, we observe that the population begins to concentrate around the $\ket{N/2,- N/2}$ states.
At longer times, $\tau_{\rm GHZ,3}\approx1.3$, a generalized GHZ state appears, with only $c_{\pm N/2}\neq0$, while all other coefficients varnish.
For all the states shown, a final $\pi/2$ rotation around the $x$-axis has been applied to make the GHZ structure more visible in Fig.~\ref{fig:2}(c)~\cite{molmer1999multiparticle}. We emphasize these are not ideal GHZ states with $c_{N/2}=c_{-N/2}=1/\sqrt{2}$, and only exist when $N = 3 + 6m$ with $m \in \mathbb{Z}^+$ and $F_Q\approx0.9N^2$~\cite{supplement}.

\begin{figure}[!t]
\centering
\includegraphics[width=0.45\textwidth]{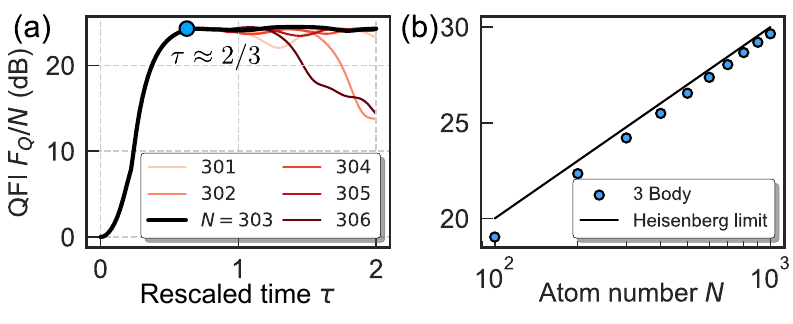}
\caption{(a) Time evolution of the QFI for atom numbers ranging from $N = 301$ to $306$. The initial rise of the QFI up to the first peak (blue dots) is well captured by the semiclassical prediction, occurring at $\tau_{\rm opt,3}\approx 2/3$, while the subsequent dynamics reveals sensitivity to the atom number.
(b) QFI evaluated at $\tau_{\rm opt,3}$ as a function of atom number. The black line denotes the HL $F_Q=N^2$.
\label{fig:3}}
\end{figure}
In Fig.~\ref{fig:3}(a), we plot the QFI as a function of time for atom numbers $N = 301$ to $306$, confirming the initial growth is universal. In Fig.~\ref{fig:3}(b), we show the QFI at $\tau_{\rm opt,3}$ versus atom number and observe values approaching the HL, $F_Q = N^2$. 
The subsequent dynamics, however, exhibit a strong dependence on the atom number modulo 6.

We compare the entanglement growth timescales with those of conventional 2BIs and find a substantially faster generation of metrologically useful states. Specifically, for 2BIs we use the standard OAT and TAT results: OAT reaches the HS QFI plateau with $F_Q = N^2/2$ at $\chi_2^{\rm OAT} t_{\rm plat,OAT}=1/\sqrt{N}$ and produces a GHZ state at $\chi_2^{\rm OAT} t_{\rm GHZ,OAT}=\pi/2$~\cite{pezze2009entanglement}, while TAT attains its optimal QFI at $\chi_2^{\rm TAT} t_{\rm opt,TAT}\sim \ln(4N)/2N$~\cite{PoggiDeutsch2023PRXQuantum}.

If the characteristic strength of 2BIs is set by $\chi_{2}$~\cite{ThompsonMomentumExchange2024,luo2025hamiltonian,luo2025realization,supplement}, a simple perturbative argument yields $\chi_3 = \chi_2 g_{\rm eff}/\Delta_c$, since 3BIs involve the exchange of an additional virtual photon at rate $g_{\rm eff}$ detuned by $\Delta_c$. However, the detunings $\Delta_{a,c}$ and the intra-cavity photon amplitude $|\alpha|$ are constrained by the requirement that adiabatic elimination remains valid for both the excited state and the cavity mode, imposing $\eta_a|\alpha|\ll 1$ and $\eta_c|\alpha|\ll 1$. Here $\eta_a=\sqrt{N}g/\Delta_{a}$ and $\eta_c=\sqrt{N}g_{\rm eff}/\Delta_{c}$, related via $\eta_c=\eta_a g/\Delta_{c}$. The 2BI strength is then $\chi_2 = g|\alpha|^2\eta_c\eta_a/N$, while the 3BIs satisfy $\chi_3 = g|\alpha|^2\eta_c^2 \eta_a / N^{\frac{3}{2}} $.

We optimize $\Delta_{a,c}$ and $|\alpha|$ under the above constraints for OAT, TAT, and 3BIs while neglecting dissipation (addressed later). Since the products $\eta_{a}|\alpha|$ and $\eta_{c}|\alpha|$ sets the characteristic timescale for 2BIs, we use the same values when evaluating 3BIs to find the ratio:
\begin{equation}
\begin{aligned}
\frac{t_{\rm GHZ,OAT}}{t_{\rm GHZ,3}} &\approx 0.3 N \eta_c \quad
\frac{t_{\rm plat,OAT}}{t_{\rm opt,3}} = \frac{3}{8} \sqrt{N}\eta_c \\
\frac{t_{\rm opt,TAT}}{t_{\rm opt,3}} &= \frac{9\ln 4N}{16} \eta_c.
\end{aligned}
\end{equation}
These comparisons show that with a reasonable choice of $\eta_c$ for 3BIs -- for example, $\eta_c = 1/2$ -- the 3BIs generate entanglement much faster than OAT and even outperform TAT for $N \gtrsim 50$. The resulting timescale is also comparable to recent proposals for creating GHZ states using all-to-all interactions~\cite{zhang2024fast,yin2025fast,ma2024quantum}.

\begin{figure}[!t]
\centering
\includegraphics[width=0.5\textwidth]{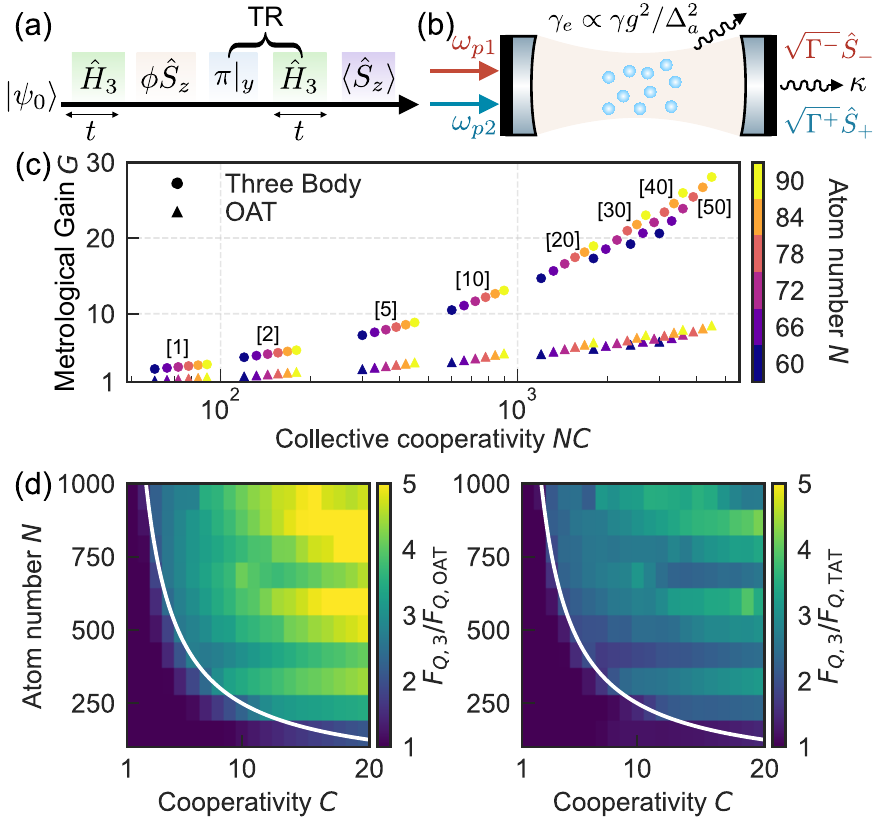}
\caption{
(a) Time-reversal (TR) protocol for quantum-enhanced sensing.
(b) Dominant dissipation channels: Spontaneous emission into free space at rate $\gamma_e$ and photon loss through cavity mirrors at rate $\kappa$, leading to two balanced collective decay modes $\hat{S}_\pm$.
(c) Metrological gain $G$ versus collective cooperativity $NC$ for 3BIs (squares) and OAT (triangles), optimized over single-particle and collective dissipation; colors denote different atom numbers, while the values in square brackets indicate the cooperativity.
(d) Ratio of the QFI for 3BIs to that of OAT (left) and TAT (right) as functions of $N$ and $C$, including collective dissipation. The QFI is evaluated at the optimal TBI evolution time.
\label{fig:4}}
\end{figure}

\textit{TR protocol--} An appealing feature of this scheme is that the rapid generation of QFI from non-Gaussian states for metrological applications can be achieved without requiring complex measurement procedures.
Interestingly in our case, the TR protocol~\cite{garttner2017measuring,garttner2018relating,lewis2019unifying} only requires a single-particle rotation along the $y$ direction of the Bloch sphere that flips $\hat{S}_{x,z}\to -\hat{S}_{x,z}$, which is illustrated in Fig.~\ref{fig:4}(a).
It begins with the preparation of an initial state, $\ket{\psi_0}$ followed by coherent evolution under the three-body Hamiltonian $\hat{U} = e^{-it\hat{H}_3}$ for a time $t$. A phase rotation $\hat{R}(\phi) = e^{-i \phi \hat{S}_z}$ is then applied, after which the dynamics are reversed using $\hat{U}^\dagger$, enabled by the echo pulse. The TR protocol can also be effectively implemented through a $\pi/3$ rotation about the $z$-axis, $\phi\to \phi+\pi/3$.

To saturate the QFI, and optimally decode the phase information, it is enough to measure the magnetization $\hat{S}_z$ instead of the fidelity. This conclusion is supported by analyzing the metrological gain relative to the standard quantum limit $\Delta \phi^2_{\rm SQL}=1/N$, $G=\Delta \phi^2_{\rm SQL}/\Delta \phi^2$, both numerically and analytically~\cite{supplement}.

\textit{Dissipative dynamics--}
In realistic implementations, both the cavity decay rate $\kappa$ and the single-particle free-space emission rate $\gamma$ from the excited state, as shown in Fig.~\ref{fig:4}(b), limit the achievable metrological gain. The key parameter determining the attainable quantum enhancement is the collective cooperativity, defined as $NC$, with $C = 4 g^2 / (\kappa \gamma)$. Previous theoretical studies incorporating both collective dephasing and single-particle spin flips in the OAT model, as well as collective decay and spin flips in the TAT and twist-and-turn models, have shown that the maximum achievable gain is constrained by $G \propto \sqrt{NC}$~\cite{chen2014cavity,davis2016approaching,norcia2018cavity,colombo2022time,leroux2010implementation,AEDGEDarkMatterProposal2020,hu2017vacuum,barberena2024fast,li2023improving}.

This scaling originates from the fact that, for the 2BIs considered here~\cite{ThompsonMomentumExchange2024,luo2025hamiltonian}, the collective dissipation rate follows $\Gamma \propto \chi_{2}\kappa / \Delta_{c}$, while the effective single-particle dissipation rate scales as $\gamma_{e} \propto \chi_{2}\Delta_{c} / (\kappa C)$. In practice, the cavity detuning $\Delta_{c}$ can be tuned to balance these two decoherence channels, thereby optimizing the achievable metrological gain. These scaling relations are independent of the atomic detuning $\Delta_{a}$, and for large atom numbers, increasing $\Delta_{a}$ further ensures that the excited state remains adiabatically eliminated.

The situation differs for the 3BIs. We find that the collective dissipation scales as $\Gamma / \chi_{3} \propto \Delta_{a} / (\gamma C)$, while the single-particle emission rate follows $\gamma_{e} / \chi_{3} \propto (\Delta_{c} / g)^{2}\Delta_{a} / (\kappa C)$, exhibiting dependence on both $\Delta_{a}$ and $\Delta_{c}$. Smaller values of $\Delta_{a}$ and $\Delta_{c}$ are thus favorable for minimizing dissipation; however, they remain constrained by adiabatic elimination conditions. Below, we present two practical scenarios in which the 3BIs offer a clear metrological gain within these constraints. See End Matter for details of the simulated models.

In the first case, we compare the optimal metrological gain of the 3BIs and OAT under the TR protocol for moderate collective cooperativity, $NC \in [60, 4500]$, taking into account both collective and single-particle dissipation. The dynamics are simulated using exact diagonalization~\cite{shammah2018open} for up to $N = 90$ atoms. We only fix $\eta_c = 1/2$ for 3BIs, while separately optimizing the cavity detuning $\Delta_c$ and the evolution time to obtain the maximal metrological gain for both models. As expected, the optimal detuning follows the scaling $\Delta_{c,{\rm opt}} \propto \sqrt{NC}$ in both cases. As shown in Fig.~\ref{fig:4}(c), 3BIs provide a pronounced enhancement over OAT within this regime. However, this advantage does not extend to arbitrarily large system sizes: for 3BIs, the atomic adiabatic elimination parameter at the optimal gain, $\eta_a = \eta_c \Delta_{c,{\rm opt}} / g \propto \sqrt{N}$, increases with $N$ and eventually breaks down the conditions.

In the second case, our goal is not to optimize the gain or its scaling, but rather to test how rapidly entanglement can be generated in a scalable regime while maintaining the validity of the adiabatic elimination.  To ensure a fair comparison, we consider a special case with same $\Delta_{a,c}$ and $|\alpha|$ for OAT, TAT, and 3BIs. To this end, we fix $\Delta_{c} = g$, which ensures $\eta_{c} = \eta_{a} \lesssim 1$, together with $|\alpha|\ll1$. We initialize the parameters with $N = 100$ and $C = 1$, choosing $\Delta_{a} = 100g$ such that $\eta_a = 0.1$. As $N$ and $C$ increase, $\eta_a$ grows as $\sqrt{NC}$, and we impose the upper bound $1/2$ for $\eta_a$ as indicated by the white line in Fig.~\ref{fig:4}(d). Below this line, the condition holds, and $\Delta_a$ is kept fixed; above it, we scale $\Delta_{a} \propto \sqrt{NC}$ to preserve the adiabatic condition with $\eta_{a} = 1/2$.
We then simulate the ratio between the QFI of the 3BIs ($F_{Q,3}$) and those of OAT ($F_{Q,{\rm OAT}}$) and TAT ($F_{Q,{\rm TAT}}$), evaluating all at the optimal evolution time of the 3BIs. Only collective dissipation is included, assuming that single-particle emission remains negligible. The results show that in the region above the white line, the 3BIs enhance the QFI by a factor of $5$ relative to OAT and by a factor of $4$ relative to TAT, demonstrating a practical metrological gain enabled by rapid entanglement generation.

It could be interesting to explore in the future the metrological performance beyond this regime by considering smaller $\Delta_c$, where the adiabatic elimination of the cavity photons breaks down, leading to even faster dynamics while requiring a full treatment of the spin–boson model.

\textit{Conclusion and outlook--}
Collective three-body interactions allow fast creation of GHZ-type entangled states with Heisenberg-limited sensitivity, far outperforming two-body schemes. This performance persists across a broad parameter range even in the presence of realistic dissipation.
While we have focused on cavity implementations, our scheme naturally extends to other boson-mediated platforms, including trapped ions, superconducting circuits, and cold molecules, where precise control of interaction strength and particle number may enable high-fidelity GHZ states. Future directions include exploring the impact of distinct decoherence channels and exploiting the system’s intrinsic $Z_3$ symmetry, which connects to bosonic codes~\cite{grimsmo2020quantum} and may inspire new error-correction strategies. More broadly, collective 3BIs expand the quantum simulation toolbox, paving the way toward the study of complex many-body models such as charge-$4e$ superconductors~\cite{berg2009charge} and lattice gauge theories~\cite{banuls2020simulating,aidelsburger2022cold}.

{\it Acknowledgments---}
We acknowledge helpful feedback on the manuscript from Diego Fallas Padilla and John Bollinger. We acknowledge  funding support from the Vannevar-Bush Faculty Fellowship,
 the National Science Foundation under Grant Numbers 1734006 (Physics Frontier Center) and  OMA-2016244 (QLCI Q-SEnSE). This work is also supported by the U.S. Department of Energy, Office of Science, National Quantum Information Science Research Centers, Quantum Systems Accelerator, 
the Heising-Simons foundation and NIST.

\bibliography{reference}

\begin{widetext}
\begin{center}
\textbf{END MATTER}
\end{center}
\end{widetext}

\textit{Realistic parameters--}
\begin{figure*}[!t]
\centering
\includegraphics[width=1\textwidth]{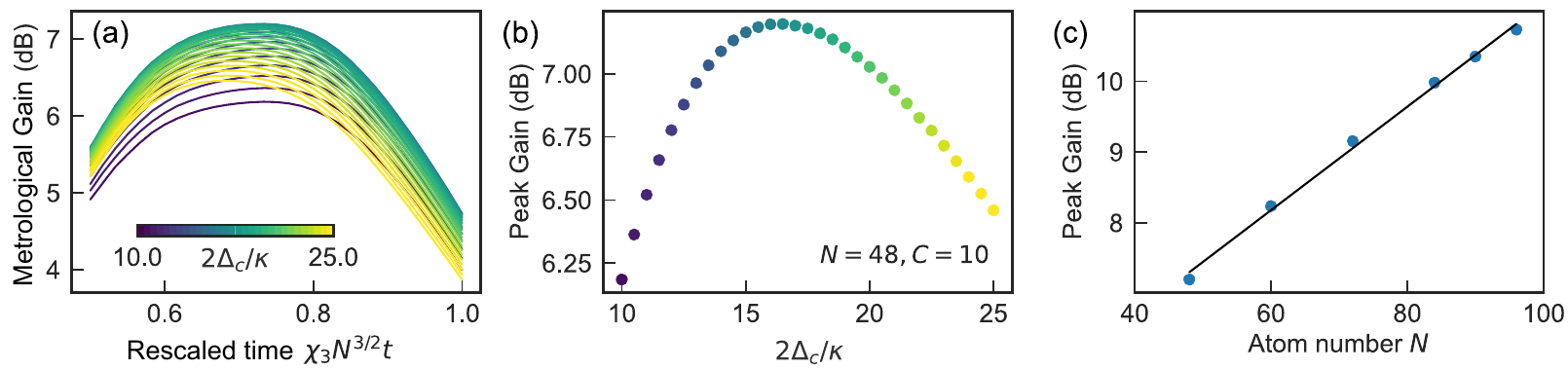}
\caption{\label{fig:5} Optimization procedure for extracting the scaling behavior.
(a) Metrological gain as a function of evolution time for fixed $N = 48$ and $C = 10$, with different colors representing various values of $d=2\Delta_c/\kappa$. For each $d$, the peak gain is identified.
(b) Peak gain as a function of $d$.
(c) Repeating the above procedure for different atom numbers $N$ reveals a linear scaling of the optimal gain with respect to $N$.} 
\end{figure*}
To numerically compare the metrological performance of OAT, TAT and 3BIs, we implement the following Hamiltonians and dissipation channels demonstrated in Refs.~\cite{luo2025realization,luo2025hamiltonian,zhang2025solitons}:
\begin{itemize}
\item 3BIs:
$\hat{H}_{3} = \chi_{3} (\hat{S}_{+}^{3} + \hat{S}_{-}^{3})$,
$\hat{L}_{3}^{+} = \sqrt{\Gamma^{+}} \hat{S}_{+}$,
$\hat{L}_{3}^{-} = \sqrt{\Gamma^{-}} \hat{S}_{-}$,
with $\chi_{3} = |\alpha|^{2} g_{\rm eff}^{3} / \Delta_{c}^{2}$ and $\Gamma^{+} = \Gamma^{-} = \Gamma$, also illustrated in Fig.~\ref{fig:4}(b).

\item OAT:
$\hat{H}_{\mathrm{OAT}} = \chi_{2}^{\mathrm{OAT}} \hat{S}_{x}^{2}$,
$\hat{L}_{2}^{\mathrm{OAT}} = \sqrt{\Gamma_{2}^{\mathrm{OAT}}} \hat{S}_{x}$,
with $\chi_{2}^{\mathrm{OAT}} = 4\chi_{2}$ and $\Gamma_{2}^{\mathrm{OAT}} = 4\Gamma$.

\item TAT:
$\hat{H}_{\mathrm{TAT}} = \chi_{2}^{\mathrm{TAT}} (\hat{S}_{x}^{2} - \hat{S}_{z}^{2})$,
$\hat{L}_{2}^{\mathrm{TAT}} = \sqrt{\Gamma_{2}^{\mathrm{TAT}}} (\tfrac{\sqrt{2}+1}{\sqrt{2}-1} \hat{S}_{+} + \hat{S}_{-})$,
with $\chi_{2}^{\mathrm{TAT}} = \tfrac{4}{3}\chi_{2}$ and $\Gamma_{2}^{\mathrm{TAT}} = (\tfrac{\sqrt{2}-1}{\sqrt{3}})^{2}\Gamma$.
\end{itemize}
Here, $\chi_{2} = |\alpha|^{2} g_{\rm eff}^{2}/\Delta_{c}$ and $\Gamma = |\alpha|^{2} g_{\rm eff}^{2} \kappa/\Delta_{c}^{2}$. We assume a total intracavity photon number of $2|\alpha|^{2}$ when determining the corresponding prefactors.
The single-particle emission rate is given by $\gamma_{e} = 2\gamma |\alpha|^{2} g^{2} / \Delta_{a}^{2}$.
We simulate the associated spin-flip processes within the $\ket{\uparrow/\downarrow}$ manifold, while~\cite{supplement} presents simulations of single-particle atom loss using the quantum trajectory method~\cite{zhang2018monte}.

With the cavity cooperativity defined as $C = 4g^{2} / (\kappa\gamma)$, we consider two parameterization strategies as discussed in the main text.
In Fig.~\ref{fig:4}(c), we first assume that the adiabatic elimination of the excited state remains valid and fix $\eta_{c} = 1/2$.
Under this condition, the relevant ratios between the dissipative rates and the strength of 3BIs are given by
\begin{equation}
\frac{\Gamma}{\chi_{3}} = \frac{\sqrt{N}}{\eta_{c}}\frac{\kappa}{\Delta_{c}}
\quad
\frac{\gamma_{e}}{\chi_{3}} = \frac{8\sqrt{N}}{\eta_{c}}\frac{\Delta_{c}}{\kappa C}.
\end{equation}
Accordingly, $\Delta_{c}$ can be optimized to balance collective and single-particle dissipation.
Although $\Gamma/\chi_{3}$ itself does not explicitly depend on $\Delta_{c}$, the parameter $\eta_{c} \propto 1/\Delta_{c}$ is fixed in our analysis, which indirectly couples $\Delta_{c}$ to the overall dissipation balance.

\begin{figure}[!t]
\includegraphics[width=0.6\columnwidth]{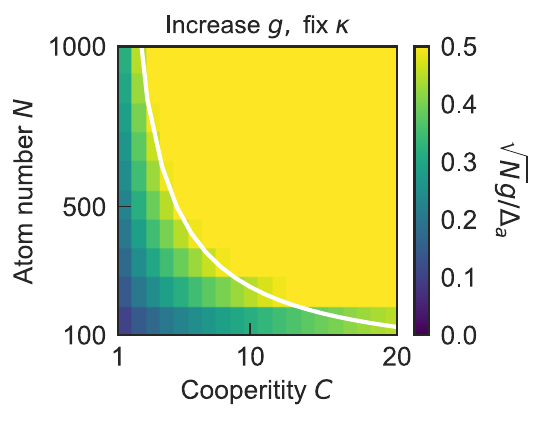}
\caption{Adiabatic elimination condition for the excited state $\eta_a=\sqrt{N} g / \Delta_a$ as a function of atom number $N$ and cavity cooperativity $C$.
To increase $C$, we scale $g \propto \sqrt{C}$ while keeping $\kappa$ and $\gamma$ fixed.
The white line marks the boundary $\eta_a = 1/2$.\label{fig:6}}
\end{figure}

The optimization procedure for Fig.~\ref{fig:4}(c) in the main text is detailed in Fig.~\ref{fig:5}. For given values of atom number $N$ and cooperativity $C$, we scan the evolution time $t$ for various values of $d=2\Delta_c/\kappa$, as shown in Fig.~\ref{fig:5}(a). For each $d$, the optimal metrological gain (versus time) is extracted and plotted in Fig.~\ref{fig:5}(b). Finally, we repeat this procedure for different values of $N$ and interestingly, observe a linear scaling with respect to $N$ even with moderate cooperativity, as shown in Fig.~\ref{fig:5}(c). However, as discussed in the main text, this fixed-$\eta_{c}$ parameterization is not further scalable with increasing atom number, motivating an alternative scheme introduced in Fig.~\ref{fig:6}.

\begin{figure}[!t]
\includegraphics[width=1\columnwidth]{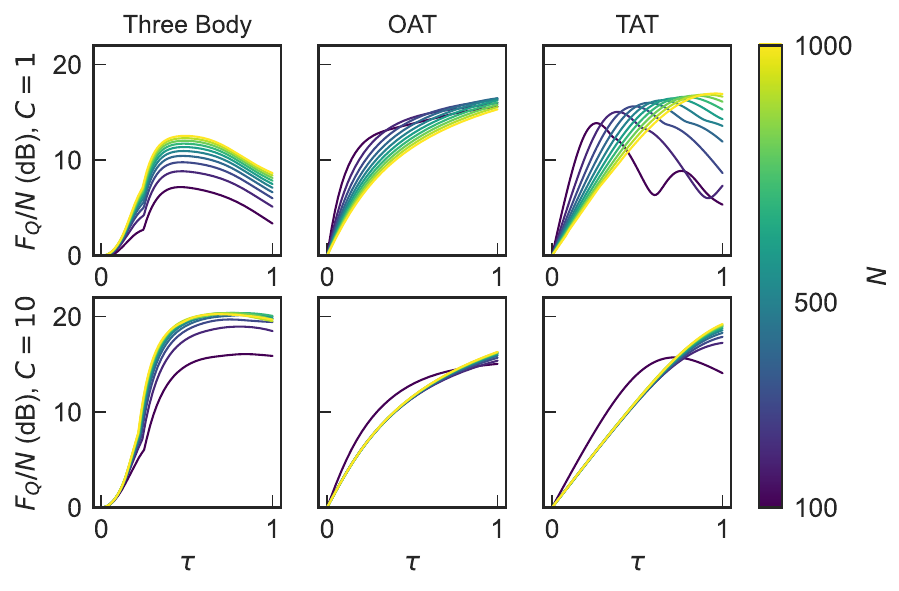}
\caption{QFI gain as a function of time for the 3BIs, OAT, and TAT. The top panel corresponds to $C = 1$, and the bottom panel to $C = 10$. Color maps correspond to atom numbers from 100 to 1000.\label{fig:7}}
\end{figure}

We introduce the dimensionless ratios 
$\beta_{\kappa} = \Delta_{a} / \kappa$ and $\beta_{\gamma} = \Delta_{a} / \gamma$, which yield $\Delta_{a}/g = 2 \sqrt{\beta_{\kappa} \beta_{\gamma} / C}$.
The ratios between the relevant rates can then be expressed as
\begin{equation}
\frac{\Gamma}{\chi_{3}} = \frac{4 \beta_{\gamma}}{C}
\quad
\frac{\gamma_{e}}{\chi_{3}} = \frac{8 \beta_{\kappa}}{C}
\quad
\frac{\chi_{2}}{\chi_{3}} = \frac{\Delta_a}{g}=2\sqrt{\frac{\beta_\kappa \beta_\gamma}{C }}.
\end{equation}
by assuming $g=\Delta_{c}$. In the numerical simulations including superradiance, we start with a small atom number $N = 100$ and cooperativity $C = 1$, and define the \textit{constants} $\beta_{\kappa} = 500$ and $\beta_{\gamma} = 5$, which give $\Delta_a / g = 100$. We then increase $N$ and $C$, testing whether the adiabatic elimination condition $\eta_a \leq 1/2$ is satisfied. 
In Fig.~\ref{fig:6}, we plot $\eta_a$ while scanning $N$ and $C$, with the contour $\eta_a = 1/2$ marked as the white line. Below this line, $\Delta_a$ is kept fixed, and we take $\Gamma / \chi_{3} = 4 \beta_{\gamma} / C$ and $\chi_{2} / \chi_{3} = 2 \sqrt{\beta_{\kappa} \beta_{\gamma} / C}$ in the simulations. Above the white line, $\Delta_a$ is increased by a factor $r = \sqrt{N C / (\beta_{\kappa} \beta_{\gamma})}$, and we choose $\Gamma / \chi_{3} = 4 r \beta_{\gamma} / C$ and $\chi_{2} / \chi_{3} = 2 \sqrt{N}$. For a given $N$ and $C$, we first determine the optimal evolution time for the 3BIs. We then evaluate the QFI of OAT and TAT at this same time to provide a direct comparison.

In Fig.~\ref{fig:7}, we plot the QFI gain as a function of time for $C = 1$ and $C = 10$. 
For $C = 1$, where $\Delta_a$ is kept fixed, the gain is found to be limited compared with OAT and TAT. 
This can be attributed to two factors: first, the dynamics is not significantly faster due to the large prefactor $\chi_{2} / \chi_{3} = 100$; and second, the generated state is more susceptible to superradiant decay. 
In contrast, for $C = 10$, above the white line, we have $\chi_{2} / \chi_{3} = 2 \sqrt{N}$ and indeed observe much faster dynamics than in OAT and TAT, accompanied by greater metrological gain. 

\end{document}


\title{Collective three-body interactions enable a robust quantum speedup: Supplemental Materials}
\author{Haoqing Zhang}
\affiliation{JILA, NIST and Department of Physics, University of Colorado, Boulder, Colorado 80309, USA}
\affiliation{Center for Theory of Quantum Matter, University of Colorado, Boulder, Colorado 80309, USA}
\author{Anjun Chu}
\affiliation{JILA, NIST and Department of Physics, University of Colorado, Boulder, Colorado 80309, USA}
\affiliation{Center for Theory of Quantum Matter, University of Colorado, Boulder, Colorado 80309, USA}
\author{Chengyi Luo}
\affiliation{JILA, NIST and Department of Physics, University of Colorado, Boulder, Colorado 80309, USA}
\author{Chitose Maruko}
\affiliation{JILA, NIST and Department of Physics, University of Colorado, Boulder, Colorado 80309, USA}
\author{Eliot A.~Bohr}
\affiliation{JILA, NIST and Department of Physics, University of Colorado, Boulder, Colorado 80309, USA}
\author{James K. Thompson}
\affiliation{JILA, NIST and Department of Physics, University of Colorado, Boulder, Colorado 80309, USA}
\author{Ana Maria Rey}
\affiliation{JILA, NIST and Department of Physics, University of Colorado, Boulder, Colorado 80309, USA}
\affiliation{Center for Theory of Quantum Matter, University of Colorado, Boulder, Colorado 80309, USA}
\date{\today}
\maketitle
\section{Multi-body interactions in cavity-QED system \label{sec:mb}}
In this section, we sketch the derivation of multi-body interactions in cavity-QED-based matter-wave interferometers~\cite{ThompsonMomentumExchange2024,luo2025realization,wilson2024entangled,zhang2025solitons,ReyZhang2023BO}, with full details provided in the supplementary material of Ref.~\cite{luo2025hamiltonian}.
As illustrated in Fig.~1 of the main text, a single ground-state level $\ket{g}$ within the atomic manifold is coupled to an excited state $\ket{e}$, separated by energy $\omega_a$,  via a single cavity mode $\hat{a}$ of frequency $\omega_c$ with finite decay rate $\kappa$.  The cavity detuning from  the $e$  to $g$ atomic transition  is  $\Delta_a = \omega_c - \omega_a$. The atom-cavity coupling is characterized by a peak single-photon Rabi frequency $2g$.

If  $\Delta_{a} \gg g \sqrt{N \langle \hat{a}^{\dagger} \hat{a} \rangle}$ and $\Delta_a \gg \gamma$, where $\gamma$ denotes the spontaneous emission rate of the excited state $\ket{e}$, we can adiabatically eliminate the excited state and  obtain an effective  atom-cavity Hamiltonian acting only on the ground state levels:
\begin{equation}
\hat{H}_S = \hat{H}_0 + g_{\rm eff}\hat{a}^{\dagger}\hat{a}\left(\hat{S}_{+}+\hat{S}_{-}\right) 
    + \left(\epsilon_{1}e^{-i\omega_{p1}t} + \epsilon_{p2}e^{-i\omega_{2}t} \right)\hat{a}^{\dagger} + \rm h.c., \label{eq:hlab}
\end{equation}
where $\hat{H}_0 = \omega_c \hat{a}^{\dagger} \hat{a} + \omega_{z} \hat{S}_{z}$ and $g_{\rm eff}=g^{2}/\Delta_{a}$. In the experiment, Bragg lasers are applied to prepare atoms in a superposition of two momentum states centered around $p_0 \pm \hbar k$, which we encode as the pseudo-spin states $\ket{\uparrow} \equiv \ket{p_0 + \hbar k}$ and $\ket{\downarrow} \equiv \ket{p_0 - \hbar k}$, with an energy splitting $\omega_z$.

Two classical dressing lasers are applied to the cavity with amplitudes $\epsilon_{1,2}$ and frequencies $\omega_{p1,2}$, satisfying the condition $\omega_{p2} - \omega_{p1} = 3\omega_z$ in the experiment. When the detunings  $\Delta_{c2} = \omega_{p2} - \omega_z - \omega_c$, and $\Delta_{c1} = \omega_{p1} + \omega_z - \omega_c$, satisfy $|\Delta_{c2,c1}|\gg g_{\rm eff}|\alpha_{1,2} |\sqrt{N}$, with
$\alpha_{1,2} = \frac{\epsilon_{1,2}}{i\kappa/2 + (\omega_{1,2} - \omega_c)}$, the cavity-built classical fields  inside the cavity, we can  adiabatically eliminate the cavity field fluctuations, and obtain an effective three-body interactions (3BIs):
\begin{equation}
\hat{H}_{3} = \chi_3 \hat{S}_+^3+ \chi_3^* \hat{S}_-^3, \quad \chi_3 \approx
\frac{g_{\rm eff}^{3}}{\Delta_{c1}\Delta_{c2}}\alpha_{1}^{*}\alpha_{2},
\end{equation} The phase of $\chi_3$ can be tuned by adjusting the relative phase between the two dressing lasers. In the following discussion, we consider a symmetric pumping configuration by setting $\alpha_1 = \alpha_2\equiv \alpha$ and $-\Delta_{c1} = \Delta_{c2}\equiv \Delta_c$, and take $\chi_3$ to be real. 

The effective Hamiltonian is also accompanied by two collective superradiant decay channels~\cite{ThompsonMomentumExchange2024}, which arise from a process in which an atom absorbs a pump photon and subsequently emits a cavity photon that leaks out of the cavity at a finite rate $\kappa$:
\begin{equation}
L_{+} = \sqrt{\Gamma^+}\hat{S}_{+}, \quad L_{-} = \sqrt{\Gamma^-}\hat{S}_{-}, \quad \Gamma^{\pm} = g_{\rm eff}^{2}|\alpha|^{2} \frac{\kappa}{\Delta_{c}^{2}}.
\end{equation}
With the two superradiant decay rates are balanced under above symmetric pumping configuration~\cite{luo2025realization,zhang2025solitons}.
Single-particle dissipation is modeled as a spin-flip process with the jump operators
\begin{equation}
\hat{L}_{s,i} = \sqrt{\gamma_e} \hat{s}_{-,i} \quad \gamma_e \approx 2 g_{\rm eff} |\alpha|^2\frac{\gamma }{\Delta_a}
\end{equation}
In Sec.~\ref {sec:loss}, we also benchmark with a more realistic single-particle atom loss processes, which account for light induced scattering into momentum states outside the the pseudo-spin states participating  in the  matter-wave interferometer.


\section{Unitary dynamics \label{sec:uni}}
\subsection{Semi-classical analysis of three-body interactions}
In this part, we estimate the time scales for the dynamics via the semi-classical phase space method, introduced in Fig.~\cite{PoggiDeutsch2023PRXQuantum}. We start with the Heisenberg equations of motion for collective spin operators, $d\langle \hat{S}_\alpha\rangle = i\langle[\hat{H}_3,\hat{S}_{\alpha}]\rangle$ with $\alpha=x,y,z$. As discussed in the main text, we define the classical variables $(x,y,z)\equiv (\langle \hat{S}_x\rangle,\langle \hat{S}_y\rangle,\langle \hat{S}_z\rangle)/(N/2)$, with the mean-field equations of motion:
\begin{equation}
\frac{d}{dt}\left(\begin{array}{c}
x\\
y\\
z
\end{array}\right)=\frac{3}{2}\chi_{3}N^{2}\left(\begin{array}{c}
-2xyz\\
z(y^2-x^2)\\
3x^{2}y-y^{3}
\end{array}\right) \label{eq:mean}.
\end{equation}
There exist six stable fixed points on the equators $(\cos\varphi,\sin\varphi,0)$ with $\varphi=n\pi/3,n\in \mathbb{Z}$, and two unstable fixed points on the north and south pole $(0,0,\pm 1)$~\cite{luo2025realization}. 
These structure are consistent with the symmetry of the Hamiltonian: a $\pi$ rotation about the $\hat{y}$ axis, $e^{i\pi \hat{S}_y} \hat{H}_3 e^{-i\pi \hat{S}_y} = -\hat{H}_3$, or a $\pi/3$ rotation about the $\hat{z}$ axis, $e^{i\pi/3 \hat{S}_z} \hat{H}_3 e^{-i\pi/3 \hat{S}_z} = -\hat{H}_3$, effectively flips the sign of $\chi_3$. In the discussion below, we focus on the case with $\chi_3 > 0$.

As discussed in the main text, the variance around the unstable point at the north pole, is evolving inwards along $\varphi = \pi/6 + 2n\pi/3$ and outwards along $\varphi = \pi/2 + 2n\pi/3$, with $n\in \mathbb{Z}$. Without loss of generality, we study the dynamics along the great circle with $x=0$, thus $y=\sqrt{1-z^2}$ to obtain:
\begin{equation}
\frac{dz}{dt} = -\frac{3}{2}\chi_3 N^2 (1-z^2)^{3/2}.
\end{equation}

To study the time for the peak Quantum Fisher information (QFI) with initial coherent spin state prepared along $\hat{z}$ axis, we apply the method in Ref.~\cite{PoggiDeutsch2023PRXQuantum}, where the point at the edge of the uncertainty patches $b_i=(0,\sqrt{1/N},\sqrt{1-1/N})$, evolves towards the equators $b_f=(0,1,0)$. Therefore, we find:
\begin{equation}
\frac{3}{2} \chi_3 N^2 t_{\rm opt, 3} = -\int_{\sqrt{1-1/N}}^{0} \frac{dz}{(1-z^2)^{3/2}} \Rightarrow t_{\rm opt,3} = \frac{2}{3 \chi_3 N^{3/2}},
\end{equation}
assuming $N \gg 1$. The estimated optimal time matches the numerical result in Fig. 2 in the main text.

\subsection{Generation of GHZ-like states}
\begin{figure}[!thb]
\centering
\includegraphics[width=0.7\textwidth]{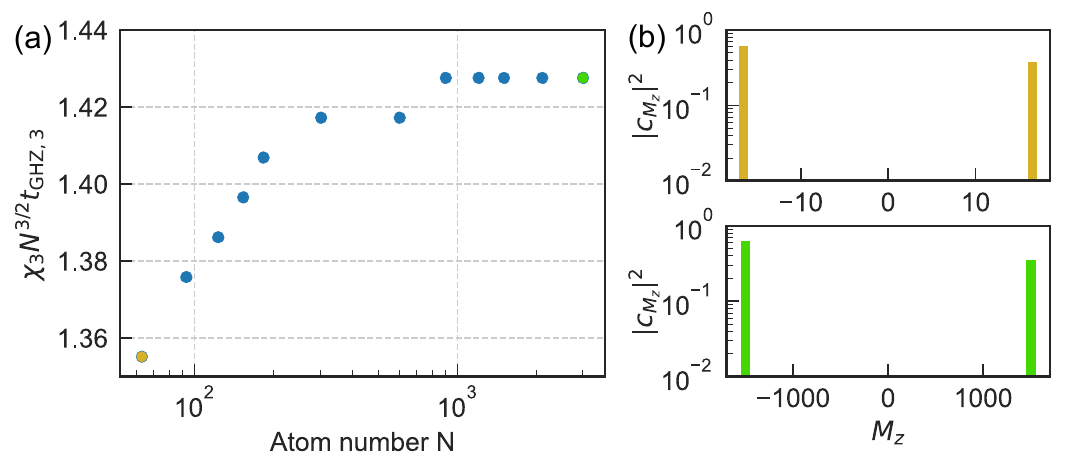}
\caption{\label{fig:GHZ} (a) Optimal time $t_{\rm GHZ,3}$ for generating a GHZ-like state as a function of atom number $N$. The scaled time $\chi_3 N^{3/2} t_{\rm GHZ,3}$ saturates to 1.42 in the large-$N$ limit.
(b) Population distribution of the GHZ-like state at the optimal time for $N=33$ (brown) and $N=3003$ (green). The populations outside the $M_z = \pm N/2$ Dicke states remain below $10^{-2}$.}
\end{figure}
In Fig.~\ref{fig:GHZ}(a), we numerically simulate the optimal time for generating a GHZ-like state as a function of the atom number, ranging from $N=33$ to $N=3003$ with $N=6n+3$, $n\in\mathbb{Z}$. We use the population $|c_{N/2}|^2 + |c_{-N/2}|^2$ as a metric to evaluate the prepared state.
We find that as the atom number increases, the optimal time $\chi_3 N^{3/2} t$ saturates to approximately 1.42.

In Fig.~\ref{fig:GHZ}(b), we show the spin distribution function for $N=33$ (brown) and $N=3003$ (green).
For both cases, we observe that $|c_{N/2}|^2 \approx 0.62$ for $N=33$ and $|c_{N/2}|^2 \approx 0.64$ for $N=3003$. Meanwhile, the remaining populations at Dicke states with $M_z \neq \pm N/2$ are below $10^{-2}$, and this behavior is consistent across all atom numbers simulated.
Note that the prepared state is not an exact GHZ state with $|c_{\pm N/2}| = 1/\sqrt{2}$; however, the state features high QFI $F_Q= 0.96 N^2$ remains useful for quantum sensing~\cite{kielinski2024ghz} and other quantum information tasks~\cite{yin2025fast}.

In Fig. 2(c) of the main text, we plot the spin Wigner distribution of the generated state $\rho_t=e^{-i\hat{H}_3t} \ket{\psi_0}\bra{\psi_0} e^{i\hat{H}_3t}$ defined as: 
\begin{equation}
W(\theta,\varphi)=\sum_{S=0}^N \sum_{M_z=-S}^{S} \mathrm{Tr}(\rho_t T^\dagger_{S, M_z}) Y_{S, M_z}(\theta,\varphi),
\end{equation} 
where $T_{S,M_z}$ are irreducible tensor operators forming a basis for spin observables, and $Y_{S,M_z}$ are standard spherical harmonics on the Bloch sphere. This representation allows us to visualize the quantum state in spin phase space, with negative regions indicating nonclassical features. It's known for a GHZ state with odd particle number, a final $\pi/2$-rotation along $\hat{x}$ axis to visualize the interference patterns~\cite{molmer1999multiparticle}, as shown in Fig. 2(c).

\subsection{Comparison with two-body interaction}
\begin{figure}[!thb]
\centering
\includegraphics[width=0.5\textwidth]{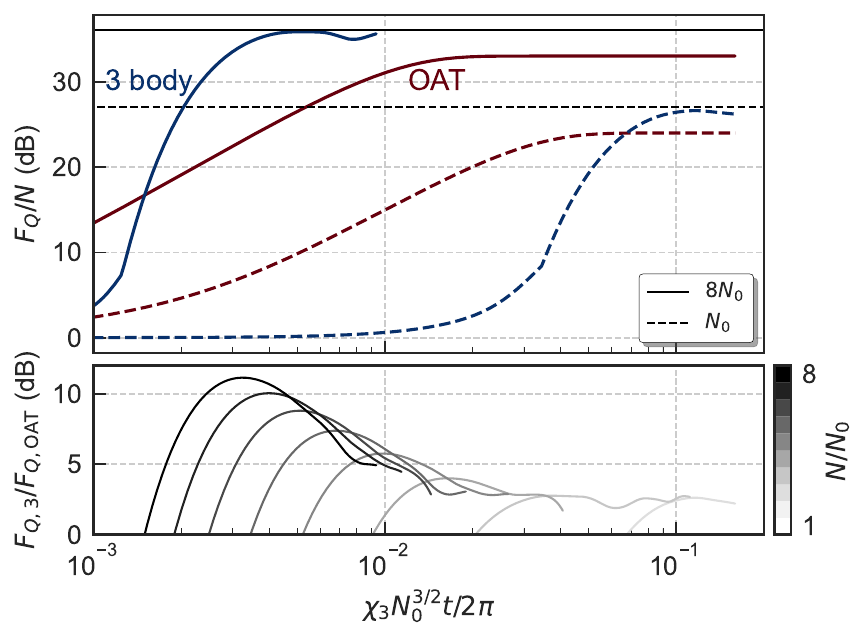}
\caption{\label{fig:oat} Upper panel: Time evolution of the QFI under 3BIs (blue) and OAT (red). Dashed lines correspond to $N=500$ atoms, while solid lines correspond to $N=4000$ atoms.
The vertical lines represent the corresponding Heisenberg limit $F_Q=N^2$.
Lower panel: Ratio of QFI between 3BIs and OAT as a function of total atom number, ranging from $N=500$ to $N=4000$. The 3BIs yield over 10 dB peak enhancement in QFI compared to the OAT scheme at $N = 4000$.
All simulations are performed with $\chi_2 = 10 \chi_3 / \sqrt{N}$.}
\end{figure}
In this section, we compare the performance of 3BIs with the well-known two-body one-axis twisting (OAT) model, or equivalently the all-to-all Ising interaction. As discussed in the main text, the two interaction strengths are related by $\chi_2 = \chi_3 \sqrt{N}/\eta$.
In Fig.~\ref{fig:oat}, we fix $\eta = 0.1$ and examine the growth of the QFI for $N = 500$ (dashed lines) and $N = 4000$ (solid lines) in the upper panel. The blue curves correspond to the QFI under 3BIs, denoted $F_{Q,3}$, while the red curves show the QFI of the OAT model, $F_{Q,\rm OAT}$. The evolution time is rescaled using a reference atom number $N_0 = 500$.
We find that the 3BIs reach the Heisenberg limit significantly faster than OAT. Moreover, as the atom number increases, $F_{Q,3}$ surpasses $F_{Q,\rm OAT}$ at progressively earlier times.
In the lower panel, we plot the ratio $F_{Q,3} / F_{Q,\rm OAT}$ as a function of time for $N$ ranging from $500$ to $4000$. For $N = 4000$, the QFI generated by 3BIs exceeds that of OAT by more than 10 dB.
These results demonstrate that 3BIs enable substantially faster generation of metrologically useful quantum states compared with OAT dynamics.

\subsection{Analytical solutions from truncated wigner approximations}
\begin{figure}[!thb]
\centering
\includegraphics[width=0.9\textwidth]{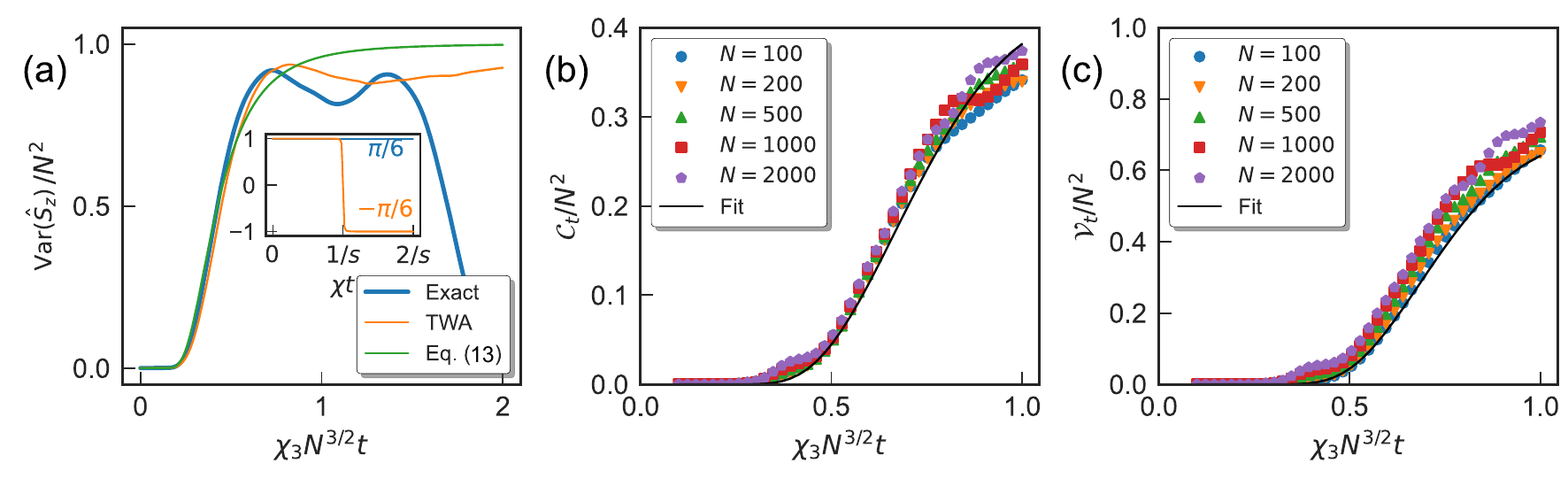}
\caption{\label{fig:twa} (a) Growth of variance $\mathrm{Var}(\hat{S}_z)$ over time for $N=500$ atoms, with the blue curve representing the exact solution, the orange curve representing the TWA result, and the green curve representing the approximated solution in Eq.~\eqref{eq:apvar}. The inset plots $z_t(s_0,\pm \pi/6)$ defined in Eq.~\eqref{eq:z} with $s_0=0.01$. 
(b), (c) Rescaled $\mathcal{C}_t$ and $\mathcal{V}_t$ as functions of time for different atom numbers. The black curves represent the fitted results in Eq.~\eqref{eq:fit}.
}
\end{figure}
In this section, we derive analytical results for the 3BIs dynamics using the semi-classical Truncated Wigner Approximation (TWA).
For a pure state, the QFI corresponds to four times the variance along the phase accumulation direction, which in our case is $4, \mathrm{Var}(\hat{S}_z)$.
Fig.~\ref{fig:twa}(a) shows the numerical results, where the TWA simulation (orange curve) and the analytical approximation from Eq.~\eqref{eq:apvar} (green curve) both agree well with the exact solution (blue curve) during the initial growth of QFI.
In the following, we detail the derivation of these results.

For the TWA simulation, we set at time $t=0$, $z_0 = 1$, while $x_0$ and $y_0$ are sampled from a Gaussian distribution $W(x_0, y_0)$ with a mean of $0$ and a variance of $1/N$.
The mean-field equations of motion in Eq.~\eqref{eq:mean} conserve the spin length,
The mean-field equations of motion in Eq.~\eqref{eq:mean} conserve the spin length $\langle \hat{S}_+\rangle \langle \hat{S}_-\rangle+ \langle \hat{S}_z\rangle^2 = (N/2)^2 + S^2$, which leads to the classical relation $x_t^2+y_t^2+z_t^2=1+s_0^2$.
They also conserve the energy,
$\langle \hat{S}_+\rangle^3 + \langle \hat{S}_-\rangle^3 = 2 S^3 \cos 3\varphi$ which gives $(x_t+iy_t)^3+(x_t-iy_t)^3=2s_0^3 \cos3\varphi_0$.
Here we define $s_0 = \sqrt{x_0^2 + y_0^2}$ and $\varphi_0 = \arctan(y_0 / x_0)$ for the initial sampled $(x_0, y_0)$.
It turns out that $s_0$ follows a Rayleigh distribution $f(s_0) = s_0 N e^{-N s_0^2 / 2}$ and $\varphi_0$ is uniformly distributed in $[0, 2\pi]$.
With the above conservation laws, we find the first-order differential equation for $z_t$:
\begin{equation}
\begin{aligned}
(\dot{z_t})^2 &= (6\chi_3 N^2)^2 [(1+s_0^2-z_t^2)^3 - s_0^6 \cos^2 3\varphi_0] \\
\Rightarrow  \mathrm{sign} (\sin 3\varphi_0) 6 \chi_3 (N/2)^2 t &=  \int_1^{z_t} \frac{dz_t}{\sqrt{(1+s_0^2-z_t^2)^3 - s_0^6 \cos^2 \varphi_0}}.
\end{aligned}
\end{equation}

In general, an analytical solution for the above integral is complex for $\cos3\varphi_0 \neq 0$. The numerical result is shown as the orange curve in Fig.~\ref{fig:twa}(a).
Instead, we can fix $\varphi_0 = \pm \pi/6$ to obtain:
\begin{equation}
\begin{aligned}
z_t(s_0,\varphi_0=\pi/6)&=\frac{1+\tilde{\chi} t(s_0^3 + s_0)}{\sqrt{1 + 2\tilde{\chi} t s_0 + \tilde{\chi}^2 t^2 (s_0^4 + s_0^2)}} \approx 1 \\
z_t(s_0,\varphi_0=-\pi/6)&=\frac{1-\tilde{\chi} t(s^{3}_0+s_0)}{\sqrt{1-2\tilde{\chi} ts_0+\tilde{\chi}^{2}t^{2}(s_0^{4}+s_0^{2})}}\approx\begin{cases}
1 & \chi t<1/s\\
-1 & \chi t>1/s 
\end{cases} \label{eq:z}
\end{aligned}  
\end{equation}
with $\tilde{\chi} = 6\chi_3 (N/2)^2$,
We plot the above two functions in the inset of Fig.~\ref{fig:twa}(a).
Then the TWA result can be approximated as follows: 
\begin{equation}
\begin{aligned}
\langle \hat{S}_z \rangle &\approx \frac{N}{2} \int_0^{\infty} ds_0 N s_0 e^{-Ns_0^2/2} \frac{z_t(s_0,\pi/6) + z_t(s_0,-\pi/6)}{2} \\
&\approx \frac{N}{4} (1+\int_0^{\frac{1}{t}} ds_0 N s_0 e^{-N s_0^2/2} - \int_{\frac{1}{ t}}^{\infty} ds_0 N s_0 e^{-N s_0^2/2}) = \frac{N}{2} (1-e^{-\frac{N}{2 (\tilde{\chi} t)^2}}) \\
\langle \hat{S}^2_z \rangle &\approx (\frac{N}{2})^2 \int_0^{\infty} ds_0 N s_0 e^{-Ns_0^2/2} (\frac{z_t(s_0,\pi/6) + z_t(s_0,-\pi/6)}{2})^2 \approx (\frac{N}{2})^2, \label{eq:apsz}
\end{aligned}
\end{equation}
with the variance is given by
\begin{equation}
\mathrm{Var}(\hat{S}_z)=(\frac{N}{2})^2 (2e^{-\frac{2}{9 (\chi_3 N^{3/2} t)^2}} - e^{-\frac{4}{9 (\chi_3 N^{3/2} t)^2}}), \label{eq:apvar}
\end{equation}
and is plotted as the blue curve in Fig.~\ref{fig:twa}(a).

\subsection{Time-reversal protocol}
\begin{figure}[!thb]
\centering
\includegraphics[width=1\textwidth]{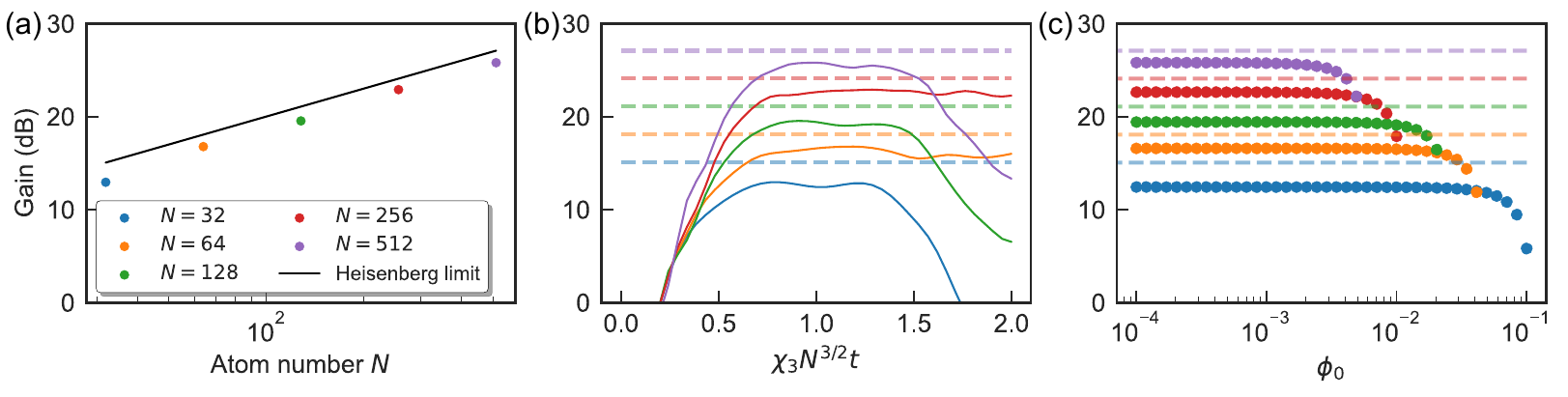}
\caption{\label{fig:tr}
Exact numerical simulation of the time-reversal protocol.  
(a) Metrological gain as a function of atom number, 
for $N = 32, \cdots, 512$ with fixed $\phi_0 = 10^{-4}$. The black line indicates the Heisenberg limit.  
(b) Metrological gain versus time for various $N$, with fixed $\phi_0 = 10^{-4}$. The dashed lines denote the Heisenberg limit for different atom numbers.  
(c) Dynamical range of the protocol, where $\phi_0$ denotes the center phase around which the signal is measured.
} 
\end{figure}
In this section, we analyze the time-reversal protocol discussed in the main text. The system is initialized in the state $\ket{\psi_0} = \ket{N/2, N/2}$, followed by phase accumulation around the $\hat{S}_z$ axis by a small angle $\phi \ll 1$. The final measurement is performed by evaluating the observable $\hat{O}$.  
The expectation value of the observable after the entire sequence is given by:
\begin{equation}
\langle \hat{O} \rangle= \bra{\psi_0} \hat{V}^\dagger \hat{O} \hat{V} \ket{\psi_0} \quad \hat{V}=e^{it \hat{H}_3} e^{-i\phi \hat{S}_z} e^{-it \hat{H}_3}.
\end{equation}
We can perform an expansion over the small angle $\phi$ to obtain:
\begin{equation}
\langle \hat{O} \rangle \approx \bra{\psi_0} \hat{O} \ket{\psi_0} + \phi^2(\bra{\psi_0} \hat{S}_z(t)\hat{O} \hat{S}_z(t) \ket{\psi_0}- \frac{1}{2} \bra{\psi_0} \hat{S}^2_z(t)\hat{O} -\hat{O}\hat{S}^2_z(t) \ket{\psi_0}),
\end{equation}
where we define $\hat{S}_z(t)=e^{i\hat{H}_3 t} \hat{S_z} e^{-i\hat{H}_3 t}$.

For a final projection measurement $\hat{P} = \ket{\psi_0}\bra{\psi_0}$ onto the initial state, which is an $N$-body operator, it is known that~\cite{lewis2019unifying,zhang2024harnessing}
\begin{equation}
\langle \hat{P} \rangle \approx 1 - \phi^2 \mathrm{Var}(\hat{S}_z) = 1 - \frac{\phi^2 F_Q}{4},
\end{equation}
where $F_Q$ denotes the QFI with respect to $\hat{S}_z$.
The corresponding phase sensitivity is given by
\begin{equation}
\Delta \phi^2 = \frac{\mathrm{Var}(\hat{P})}{\left(\partial_{\phi} \langle \hat{P} \rangle \right)^2} \approx \frac{\phi^2 F_Q / 4}{\left( \phi F_Q / 2 \right)^2} = \frac{1}{F_Q},
\end{equation}
where the variance is approximated by $\mathrm{Var}(\hat{P}) \approx \langle \hat{P} \rangle \left(1 - \langle \hat{P} \rangle \right) \approx \phi^2 F_Q / 4$ for small $\phi$.
Therefore, the projection measurement of $\hat{P}$ saturates the QFI bound.

Instead, if we measure $\hat{S}_z$, the signal now is given by:
\begin{equation}
\langle \hat{S}_z \rangle \approx \frac{N}{2}(1 - \phi^2 \mathcal{C}_t)\quad \mathcal{C}_t= \bra{\psi_0} \hat{S}^2_z(t) \ket{\psi_0} - \frac{2}{N} \bra{\psi_0} \hat{S}_z(t) \hat{S}_z  \hat{S}_z(t) \ket{\psi_0}, \label{eq:C}
\end{equation}
and for the variance
\begin{equation}
\mathrm{Var}(\hat{S}_z) \approx (\frac{N}{2})^2 \mathcal{V}_t \phi^2\quad \mathcal{V}_t = 2 \mathcal{C}_t^2 - \bra{\psi_0} \hat{S}^2_z(t) \ket{\psi_0} + (\frac{2}{N})^2\bra{\psi_0} \hat{S}_z(t) \hat{S}^2_z  \hat{S}_z(t) \ket{\psi_0}.
\end{equation}
In Fig.~\ref{fig:twa}(b) and (c), we plot the rescaled quantities $\mathcal{C}_t/N^2$ and $\mathcal{V}_t/N^2$ as a function of time for different atom numbers. We observe that all curves collapse onto a single trajectory, indicating a universal behavior.
Motivated by Eq.~\eqref{eq:apvar}, we propose the following simple ansatz
$C\left(e^{-\frac{a}{(\chi_3 N^{3/2}t)^2}} - e^{-\frac{b}{(\chi_3 N^{3/2}t)^2}}\right)$,
and fit the numerical results to obtain:
\begin{equation}
\mathcal{C}_t = N^2 \left(e^{-\frac{7}{9(\chi_3 N^{3/2}t)^2}} - e^{-\frac{23}{9(\chi_3 N^{3/2}t)^2}}\right), \quad
\mathcal{V}_t = 4 N^2 \left(e^{-\frac{10}{9(\chi_3 N^{3/2}t)^2}} - e^{-\frac{16}{9(\chi_3 N^{3/2}t)^2}}\right), \label{eq:fit}
\end{equation}
as shown by the black curves.
As illustrated in Fig.~\ref{fig:twa}(b) and (c), these expressions provide accurate approximations for $\chi_3 N^{3/2}t \lesssim 1$.

Additionally, since $\mathcal{C}_t \propto N^2$, the condition $\phi \lesssim 1/N$ ensures that the protocol operates within the dynamical range where only lower-order terms contribute significantly.  
However, it is also necessary to choose a nonzero center phase to ensure a finite signal derivative and non-vanishing variance~\cite{zhang2024harnessing,ma2024quantum}.  
In the following, we denote such a center phase by $\phi_0$.
The metrological gain with respect to the standard quantum limit $\Delta \phi^2_{\rm sql} = 1/N$ can be calculated as:
\begin{equation}
G = \frac{(\partial_{\phi}\langle \hat{S}_z \rangle)^2}{N\, \mathrm{Var}(\hat{S}_z)} \approx \frac{4 \mathcal{C}_t^2}{N\, \mathcal{V}_t} \approx
N \frac{\left(e^{-\frac{7}{9(\chi_3 N^{3/2}t)^2}} - e^{-\frac{23}{9(\chi_3 N^{3/2}t)^2}}\right)^2}
{e^{-\frac{10}{9(\chi_3 N^{3/2}t)^2}} - e^{-\frac{16}{9(\chi_3 N^{3/2}t)^2}}}, \label{eq:ana}
\end{equation}
In Fig.~\ref{fig:tr}(a), we plot the metrological gain versus atom numbers, showing performance close to the Heisenberg limit for a fixed center phase $\phi_0 = 10^{-4}$.  
In Fig.~\ref{fig:tr}(b), we present the gain as a function of time for various atom numbers, also with $\phi_0 = 10^{-4}$.  
Figure~\ref{fig:tr}(c) shows the optimal gain (maximized over time) as a function of $\phi_0$, revealing a plateau that indicates the gain is nearly independent of the center phase for $\phi_0 \lesssim 1/N$, consistent with the analytical expression in Eq.~\eqref{eq:ana}.

\section{Dissipative dynamics}

\subsection{Numerical simulations \label{sec:num}}

For the numerical simulations with single-particle spin flips, we use the exact permutationally invariant subspace method developed in Ref.~\cite{shammah2018open} to compute $\mathrm{Var}(\hat{S}_z)$ and the derivative $\partial \langle \hat{S}_z \rangle/\partial\phi$.
To improve the accuracy of the derivative in the presence of dissipation, we rewrite the signal as
\begin{equation}
\langle \hat{S}_z \rangle = \mathrm{Tr} \left( \hat{S}_z e^{\mathcal{L}_{-\chi_3} t} \left( e^{i\phi_0 \hat{S}_z} e^{\mathcal{L}_{\chi_3} t}(\ket{\psi_0} \bra{\psi_0}) e^{i\phi_0 \hat{S}_z} \right) \right),
\end{equation}
where $\mathcal{L}_{\chi}$ denotes the Lindblad generator with 3BI strength $\chi$, including both single-particle spin flips and balanced superradiance.
From this form, we observe that the derivative of the signal with respect to $\phi$ is given by the imaginary part of a two-point correlation function:
\begin{equation}
\frac{\partial}{\partial\phi}\langle \hat{S}_z \rangle = -i \left[ \mathrm{Tr} \left(\hat{S}_z e^{\mathcal{L}_{-\chi_3} t} (\hat{S}_z \rho_t) \right) - \mathrm{Tr} \left( \hat{S}_z e^{\mathcal{L}_{-\chi_3} t} (\rho_t \hat{S}_z) \right) \right] = 2 \mathrm{Im} \langle \hat{S}_z(2t) \hat{S}_z(t) \rangle, \label{eq:two}
\end{equation}
where we now define $\rho_t = e^{i\phi_0 \hat{S}z} e^{\mathcal{L}{\chi_3} t}(\ket{\psi_0} \bra{\psi_0}) e^{i\phi_0 \hat{S}_z}$. Throughout all numerical simulations, we evaluate the derivative using Eq.~\eqref{eq:two}.


\subsection{Atom loss \label{sec:loss}}
\begin{figure}[!thb]
\centering
\includegraphics[width=1\textwidth]{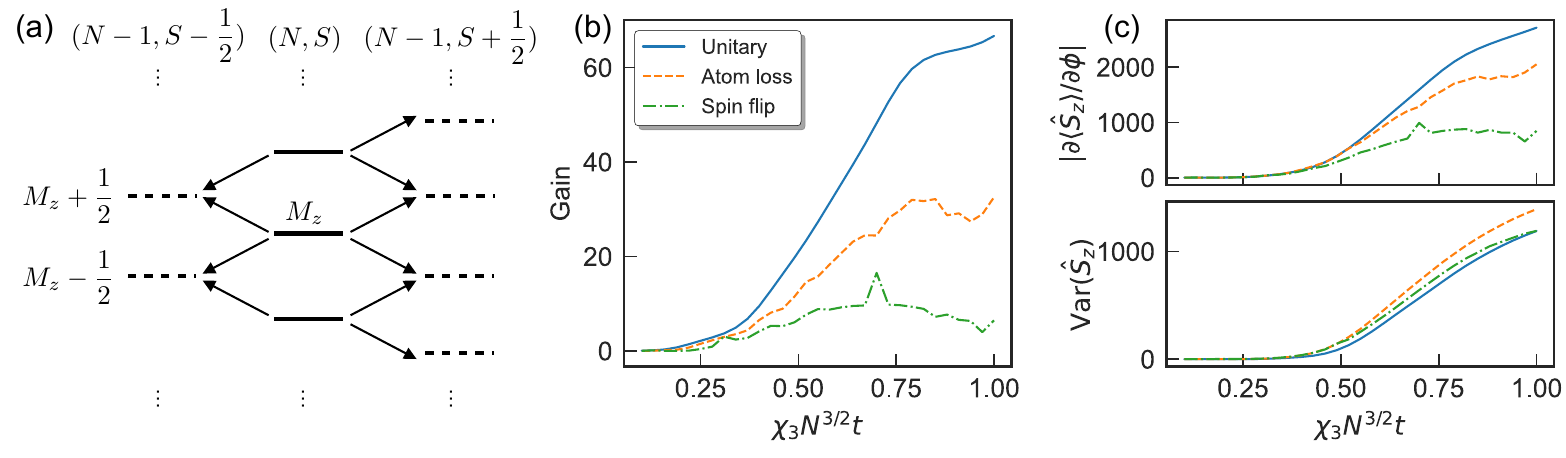}
\caption{\label{fig:loss} Monte Carlo simulations of the dynamics of $N = 93$ atoms and $C=10$ with atom loss processes.
(a) Illustration of quantum jumps in the Dicke state manifolds due to atom loss. Each loss event couples the state $(N, S, M_z)$ (solid levels) to $(N-1, S \pm \frac{1}{2}, M_z \pm \frac{1}{2})$ (dashed levels), with transition amplitudes given in Ref.~\cite{zhang2018monte}.
(b) Metrological gain vs evolution time for $\phi_0 = 1/N$. The orange dashed line shows the result with atom loss via the quantum jump method, while the blue solid and green dotted lines represent unitary evolution and with spin flip, respectively, both computed using ED.
(c) Derivative and variance of the signal vs evolution time.
}
\end{figure}
In this section, we incorporate atom loss, the dominant single-particle dissipation in matter-wave systems~\cite{wilson2024entangled}. This process corresponds to leakage from the pseudo-spin manifold into other momentum or internal states, and is modeled using jump operators $\hat{L}_i = \sqrt{\gamma_e} \hat{l}_i$ for the loss of atom $i$.

Between quantum jumps, each trajectory evolves under a non-Hermitian Hamiltonian,
$\hat{H}_{\rm NH} = \hat{H}_3 - i \Gamma/2 (\hat{S}_+ \hat{S}_- + \hat{S}_- \hat{S}_+) - i\gamma_e/2 \sum_i \hat{l}_i^\dagger \hat{l}_i$,
until a quantum jump occurs, determined by a random number. We only need to consider a fixed total atom number $N$ and spin length $S$, thus reducing the memory requirement from $O(N^3)$ to $O(N)$~\cite{zhang2018monte}.
When a quantum jump occurs, the state originally at $(N, S, M_z)$ is coupled to $(N-1, S \pm \tfrac{1}{2}, M_z \pm \tfrac{1}{2})$ as shown in Fig.~\ref{fig:loss}(a), with transition coefficient calculated in~\cite{zhang2018monte}. Average over many trajectories provides an estimate of the expectation values of observables.

In Fig.~\ref{fig:loss}(c), we show the time evolution of the derivative $\langle \hat{S}_z \rangle$ and the variance $\mathrm{Var}(\hat{S}_z)$ for $N = 93$, cooperativity $C = 10$, central phase shift $\phi_0 = 1/N$, and detuning $2\Delta_c/\kappa = \sqrt{NC/2}$. We find that single-particle atom loss leads to a larger amplitude in the derivative compared to spin-flip dissipation. Although atom loss slightly increases the variance, the resulting metrological gain remains higher, as shown in Fig.~\ref{fig:loss}(b). A more systematic study of the impact of atom loss on metrological performance is left for future work.

\bibliography{reference}